# Java Source-code Clustering: Unifying Syntactic and Semantic Features

Janardan Misra, Vikrant Kaulgud, Gary Titus, Annervaz KM, Shubhashis Sengupta
Accenture Labs, Bangalore, India
{janardan.misra,vikrant.kaulgud,gary.titus,annervaz.k.m,shubhashis.sengupta}@accenture.com

## ABSTRACT

This is a companion draft to paper "Software Clustering: Unifying Syntactic and Semantic Features", in proceedings of the *19th Working Conference on Reverse Engineering (WCRE 2012)*. It discusses the clustering process in detail, which appeared in the paper in an abridged form. It also contains certain additional process steps which were not covered in the WCRE paper. The clustering process is described for applications with Java source-code. However, as argued in the WCRE paper, it can be seamlessly adapted to many other programming paradigms.

## 1. INTRODUCTION

Functional component architecture is an important artifact used during software design, maintenance, as well as program/code comprehension. For large systems, which have been evolving over long periods, keeping architectural documentation is generally difficult and in absence of well specified policies, it tends to drift and become obsolete. Automated discovery of component architecture from the source code is therefore an important approach for program comprehension and maintenance. Software clustering, which groups together elements of code as a component, is an important technique for extracting the high level component architecture from the underlying source code [9].

In this work, we propose an approach to extract functional component architecture from the underlying source code using a combination of both syntactic and semantic features. The proposed technique uses textual features extracted from code comments and identifiers, programming paradigm based features (e.g., inheritance relations and packaging structure in case of OO code), and dependency based features (e.g., method calls). To improve the quality of the generated clusters, following the analysis presented in [10], we apply automated weighing on the extracted features to enhance their information quality and to reduce noise. We define suitable measures to estimate relatedness between code elements in terms of the individual features and use a combination of these measures to define combined relatedness between code elements. Functional components are extracted as clusters of code elements. Similar to the graph partitioning based approaches to clustering (e.g., [4]); we treat the problem of clustering as a search problem in the space of partitions of a weighted graph. A global quality constraint is used to select partitions (clusters). We discuss methods for automated labeling of the extracted components using dominant concepts and generating inter-cluster interactions. We further discuss how the suggested approach extends to clustering at multiple hierarchical levels, to application portfolios, and for the purpose of improving precision for the query to code mapping or feature location problem.

# 2. CLUSTERING PROCESS

## Inputs

- Java Source code and byte code
- [Optional] User inputs for identifying packages and classes performing Data Access, Presentation, Models, and Utils (both technical and application specific).
- [Optional] User given textual description of functional entities.
- [Optional] User query.

## Outputs

- Cluster hierarchy: At the first level of hierarchy each cluster is a set of Classes. At the higher levels, a cluster consists of set of clusters from just one level down in the hierarchy.
- Cluster interfaces: For a Cluster of classes, a cluster interface is a list of methods by which other clusters can invoke the services offered by the classes within the cluster.
- Inter Cluster dependencies: Cluster to cluster dependency describes the set of methods by which these clusters interact with each other.
- Auto labelling for Clusters.
- Functional Entity to Cluster mapping.
- Query to Code (Class level) Mapping.

## 2.1 Overview

Clustering process starts with the step of selecting the language profile, which will be used to guide the choice of features to be extracted from the source code and the code elements to be clustered to define the functional components. Once the language profile is fixed (e.g., Java) and specified features have been extracted from the source code, we need to measure the relatedness between code elements using these features. Code elements and the combined relatedness measure between them are represented as a weighted graph, which is used as an input to the clustering process. The problem of clustering is solved by treating it as a search problem, whereby we search for the partition of the graph of code elements which maximizes a global quality constraint. We discuss heuristics to generate seed clusters and a process of local neighborhood search to select the candidate partition. Based upon the composition of the clusters, we define the interfaces of these clusters, which will guide a user in understanding which services are being offered by these clusters and their interactions with each other in terms of these services within the application. Next, we discuss how these clusters can be auto labeled so that the linguistic concepts which are built within these clusters become apparent and can give informal business semantics to these clusters.

## Step [0]: Scoping

Step [0.1]: Based upon the user inputs/built-in heuristics/combination of heuristics and user inputs, determine the boundary for the component identification by excluding those packages and classes, which are identified into presentation layer (UI), or data access layer (DA), or as Models, or Utilities.

In the following steps only remaining classes, which form the business layer of the application would be considered for clustering purposes.

# Step [1]: Feature Extraction

Inputs: Source code and byte code.

Outputs: Set of extracted features for each class and inter class structural dependencies.

Procedure:

Step[1.1] (Textual Feature Extraction)
*/* Extracting IR tokens from code comments and identifiers*/*

For each class file:
[1.1.1] Extract the code comment strings and identifiers for the public variables from the Java source code.
[1.1.2] Extract the list of words from code comment strings and variable identifiers by splitting the strings into separate words and by applying tokenization on each word.
For example, string "This ControllerClass will schedule processes" will be split into words {"This", "ControllerClass", "will", "schedule", "processes"} and tokenizer will extract tokens {"This", "Controller", "Class", "will", "schedule", "processes"}.
[1.1.3] Remove the reserved words from the lists generated in the Step[1.1.2].
[1.1.4] Remove the Stop words from the lists.
[1.1.5] Apply word spamming to bring the words to their roots. For word stemming, we use the software specific stemmer made available by A. Wiese et al. [1], which is based upon an analysis of more than 9000 open source Java projects.

Step[1.2] (Code Feature Extraction)
*/* Extracting concept words embedded in the Class names */*

For each class file:
[1.2.1] Extract the strings representing the Class name.
[1.2.2] Apply string tokenizer to extract list of "concepts words" from the string extracted in step [1.2.1].
*/* Extracting concept words embedded in the public method identifiers */*

For each Java class file:
[1.2.3] Extract a list of strings consisting of identifier of the public methods used in the class.
[1.2.4] Apply string tokenizer to extract list of "concepts words" from each string extracted in step [1.2.3].

*/* Extracting Packaging information for each class */*

For each Java class file:
[1.2.5] Extract its packaging hierarchy as a string. Extract substring describing each individual hierarchy saperated by dot ('.').

For example, if for a Class, its packaging is "com.atl.application.controlManager", it will be broken into the list of strings {"com", "atl", "application", "control", "Manager"}.

*/* Extracting inheritance relationships for each class */*

For each Java class file:
[1.2.7] Collect the list of class-names or interfaces, which are being extended or implemented by the Class. This would be indicated in the source code by the reserved words like *extends* and *implements*. This lists will be called inheritance list.

For example, if a Class definition is Class ClientAnalytics *implements* Business, Analytics, Client{ … }, the inheritance list for the class ClientAnalytics would be { Business, Analytics, Client}.

Step[1.3] (Structural Dependency based Feature Extraction)

[1.3.1] Populate dependency graph $G_{dep}$ from the byte code.
The dependency graph $G_{dep}$ has nodes representing classes. Each edge between two nodes (say node A and node B) represents that there is a method call in the source code of Class A, where a (public) method of Class B is called. Each edge contains information regarding the input and output parameter types for the called method. There may be multiple directed edges between any two nodes for different methods getting called. However, as argued in [2], in absence of detailed run time profiling of the application, it is sufficient to consider multiple instances of same method call only once without affecting the quality of clustering.

## Step [2]: Feature Analysis: Similarity Estimation

*/* To calculate Class to Class similarity scores based upon the features extracted in step [1]. */*

Step[2.1]: (Vector Space model based estimation of Textual Similarity)

[2.1.1] (Populate Co-occurrence Matrix)

i. Let D = <$Class_1$, $Class_2$, …, $Class_d$> be the sequence of Classes in the Java source. *d* is the *total* number of Classes in the source code.
ii. Let T be the sequence of all unique IR tokens occurring across the Classes.
T is the union of all the IR tokens extracted in the step [1.1].
Let *t* be the number of IR tokens in T.
iii. Create **co-occurrence** matrix C of size $d \times t$ such that
C[i, j] = frequency of occurrence of j-th IR token (from T) in the i-th Class (from D).

Rows of C (denoted as, C[1], C[2], …, C[d] ) are known as document vectors. Each document vector C[i] corresponds to the i-th class in D and measures the frequency of occurrence of various IR tokens in that class.

[2.1.2] (Apply tf-idf based automated weighting)
For each $i \in [1..d]$ and $j \in [1..t]$:
$C[i,j] \leftarrow C[i,j] \times \ln\left(\frac{d}{n}\right)$, where
**d** is total number of documents (i.e., classes) under consideration, and
**n** is the number of documents (i.e., classes) where the j-th IR token appears.
**Factor** $\ln\left(\frac{d}{n}\right)$ is referred as a IDF (inverse document frequency).
[2.1.3] (Calculate Cosine similarity [3] between each pair of Classes)

For each $i, j$ *∊ [1..d]and i ≤j:* Let

$C[i]$ and $C[j]$ be the document vectors from the co-occurrence matrix C corresponding to the i-th and j-th Classes.

Their textual similarity

$$\delta_{textual}[i,j] = \frac{\sum_{r=1}^{r=t} C[i,r]C[j,r]}{\sqrt{\sum_{r=1}^{r=t} C[i,r]^2}\sqrt{\sum_{r=1}^{r=t} C[j,r]^2}}$$

<u>Step[2.2]: (Estimation of Class name Similarity)</u>

<u>[2.2.1] (</u>Populate Class name Matrix)

i. Let $Z_c$ be the sequence of all unique “word concepts” appearing in any of the lists generated in the Step[1.2.2].

$Z_c$ is the set-union of all the word concepts extracted in the step [1.2.2].

Let $z_c$ be the *total* number of word concepts in $Z_c$.

ii. Create Class name matrix $W_{class}$ of size $d \times z_c$ such that

$W_{class}[i,j]$ = frequency of occurrence of j-th word concept (in $Z_c$) for the i-th Class (in D).

Rows of $W_{class}$ (denoted as, $W_{class}[1]$, $W_{class}[2]$, …, $W_{class}[d]$) correspond to the classes in D and measure the frequency of occurrence of various word concepts in that class name.

[2.2.2] (Apply tf-idf based automated weighting)

For each $i$ *∊ [1..d]and* $j$ *∊ [1..$z_c$]:*

$W_{class}[i,j] \leftarrow W_{class}[i,j] \times \ln\left(\frac{d}{n}\right)$, where

**d** is total number of classes in the application, and

**n** is the number of rows in $W_{class}$, where $W_{class}$[-,j] > 0.

[2.2.3] (Calculate Class name similarity between each pair of Classes)

For each $i, j$ *∊ [1..d]and i ≤j:*

Let $W_{class}[i]$ and $W_{class}[j]$ be the row vectors from $W_{class}$ corresponding to the i-th and j-th Classes in D. Their similarity is measured as as a variant of the Jaccard Similarity [3]:

$$\delta_{class}[i,j] = \frac{\sum_{r=1}^{r=z_c} min\{W_{class}[i,r], W_{class}[j,r]\}}{\sum_{r=1}^{r=z_c} max\{W_{class}[i,r], W_{class}[j,r]\}}$$

<u>Step[2.3]: (Estimation of Method name Similarity)</u>

[2.3.1] (Populate Method name Matrix)

i. Let $Z_m$ be the sequence of all unique “word concepts” appearing in any of the lists generated in the Step[1.2.4].

$Z_m$ is the set-union of all the word concepts extracted in the step [1.2.4].

Let $z_m$ be the *total* number of word concepts in $Z_m$.

ii. Create method name matrix $W_{method}$ of size $d \times z_m$ such that

$W_{method}[i,j]$ = frequency of occurrence of j-th word concept (in $Z_m$) for the i-th Class (in D).

Rows of $W_{method}$ (denoted as, $W_{method}[1]$, $W_{class}[2]$, …, $W_{method}[d]$) correspond to the classes in D and measure the frequency of occurrence of various word concepts embedded in the public methods of the class.

[2.3.2] (Apply tf-idf based automated weighting)

For each $i$ *∊ [1..d]and* $j$ *∊ [1..$z_c$]:*

$W_{method}[i,j] \leftarrow W_{method}[i,j] \times \ln\left(\frac{d}{n}\right)$, where

**d** is total number of classes in the application, and

**n** is the number of rows in $W_{method}$, where $W_{method}[-,j]$> 0.

[2.3.3] (Calculate method name similarity between each pair of Classes)

For each $i, j \in [1..d]$ and $i \leq j$:

Let $W_{method}[i]$ and $W_{method}[j]$ be the row vectors from $W_{method}$ corresponding to the i-th and j-th Java Classes in D. Their similarity is measured as:

$$\delta_{method}[i,j] = \frac{\sum_{r=1}^{r=m} min\{W_{method}[i,r], W_{method}[j,r]\}}{\sum_{r=1}^{r=z_m} max\{W_{method}[i,r], W_{method}[j,r]\}}$$

<u>Step[2.4]: (Estimation of Packaging based Similarity)</u>

[2.4.1] (Calculate Packaging similarity between each pair of Classes)
For each $i, j \in [1..d]$ and $i \leq j$:

Let $W_{package}[i]$ and $W_{package}[j]$ be the lists corresponding to the i-th and j-th Java Classes in D as generated in the Step[1.25].
Then packaging similarity between these classes is measured as a Jaccard Coefficient:

$$\delta_{package}[i,j] = \frac{|W_{package}[i] \cap W_{package}[j]|}{|W_{package}[i] \cup W_{package}[j]|}$$

Where |..| measures the size of the set.

<u>Step[2.5]: (Estimation of Inheritance based Similarity)</u>

For each class in D:
[2.5.1] <u>(Reflexivity)</u> Include the class name in its inheritance list, which was generated in the step [1.2.7].
[2.5.2] (<u>Symmetry)</u> Include the names of all those classes in its inheritance list, which have the current class in their inheritance lists.
[2.5.3] (Calculate inheritance based similarity between each pair of Classes)
For each $i, j \in [1..d]$ and $i \leq j$: Let
$W_{in}[i]$ and $W_{in}[j]$ be the inheritance lists corresponding to the i-th and j-th Classes in D. Then inheritance similarity between these classes is measured as:

$$\delta_{in}[i,j] = \frac{|W_{in}[i] \cap W_{in}[j]|}{|W_{in}[i] \cup W_{in}[j]|}$$

<u>Step[2.6]: (Estimation of Structural Similarity)</u>

[2.6.1] (Collapsing edges with same method name)
For each pair of vertices $(u, v) \in G_{dep}$:

i. Merge all the outbound edges of *u* (i.e., edges starting from vertex *u* and ending at vertex *v*) having same method name as their attribute to a single outbound edge from vertex *u* to vertex *v* with 1st attribute as the number of edges which were merged and 2nd attribute as the method name.
ii. Repeat (i) for the inbound edges of vertex *u*.
iii. Repeat (i) and (ii) for the vertex *v*.

[2.6.2] (Automated weighing)
For each vertex $u \in G_{dep}$:
For each inbound edge of *u:*
Divide its edge count (i.e., 2nd attribute) with the sum of the edge counts of all those edges having the same method name.
[2.6.3] (Collapsing edges between each pair of vertices)
For each pair of vertices $(u, v) \in G_{dep}$:
Merge all the outbound edges of *u* and *v* to a single undirected edge between *u* and *v* with only following attribute kept for the merged edge:

$$\frac{summation\ of\ all\ the\ 1^{st}\ attributes\ on\ these\ edges}{Number\ of\ (public) methods\ in\ both\ the\ classes}$$

Division by the total number of (public) methods in classes corresponding to nodes *u* and *v*, is to make the resultant similarity estimation independent of the class size.

[2.6.4] (Weight normalization)

i. Find the maximum weight of the edges in $G_{dep}$. Let it be denoted by μ.
ii. For each edge $e \in G_{dep}$:
    Divide its weight by μ.

[2.6.5] (Structural similarity)

For each *i, j ∈ [1..d]and i ≤j:*

The structural similarity between the i-th and the j-th classes is

$$\delta_{structural}[i,j] = weight\ of\ the\ edge\ in\ G_{dep}\ between\ vertices\ corresponding\ to\ the\ i^{th}\ and\ j^{th} classes$$

<u>Step[2.7]: (Estimation of Combined Similarity between each pair of classes)</u>

[2.8.1] Choose the relative significance factor for each of the similarity measures estimated in previous steps such that their sum equals 1.

Let

$\alpha_{textual}$: Significance factor for textual similarity. Default: 0.1
$\alpha_{class}$: Significance factor for class name based similarity. Default: 0.2
$\alpha_{method}$: Significance factor for method name based similarity. Default: 0.1
$\alpha_{packaging}$: Significance factor for packaging based similarity. Default: 0.2
$\alpha_{inheritence}$: *S*ignificance factor for inheritance based similarity. Default: 0.2
$\alpha_{structural}$: Significance factor for structural similarity. Default: 0.2

Such that

$$\alpha_{textual} + \alpha_{class} + \alpha_{method} + \alpha_{packaging} + \alpha_{inheritance} + \alpha_{structural} = 1$$

[2.7.2] (Combined similarity)

For each *i, j ∈ [1..d]and i ≤j:*

The combined similarity between the i-th and the j-th classes is

$$\boldsymbol{\delta_{Combined}[i,j] = \alpha_{textual} \times \delta_{textual}[i,j] + \alpha_{class} \times \delta_{class}[i,j] + \alpha_{method} \times \delta_{method}[i,j] + \alpha_{packaging} \times \delta_{packaging}[i,j] + \alpha_{inheritance} \times \delta_{inheritance}[i,j] + \alpha_{structural} \times \delta_{structural}[i,j]}$$

This yields an extended dependence Graph $\boldsymbol{G_{ext_dep}} = (\boldsymbol{V}, \boldsymbol{E}, \boldsymbol{weight})$ . Nodes in $\boldsymbol{V}$ represent Classes and edges (say, $e = (\boldsymbol{u}, \boldsymbol{v})$ in $\boldsymbol{E} \subseteq \boldsymbol{V} \times \boldsymbol{V}$) with weight $\boldsymbol{weight}(\boldsymbol{e})$ between nodes represent the combined similarity between the pair of Classes.

*Note: A high level approach to refine significance factors based upon statistical features of the source code is given in the <u>Appendix [A].</u>*

# Step [3]: Clustering

To generate clusters from$\boldsymbol{G_{ext_{dep}}}$,we define a *modularization quality criterion* (MQC), which is adapted from [4] and [5]. Aim is to select that partition of $\boldsymbol{G_{ext_dep}}$, which maximizes the cohesion within the clusters and minimizes the coupling across clusters. To define MQC, let us consider a Clustering (or partition) $P = \{C_1, C_2, \cdots, C_n\}$, where each cluster $C_i$ is a subgraph of $\boldsymbol{G_{ext_dep}}$, such that no two clusters share any nodes and clusters in P cover $\boldsymbol{G_{ext_dep}}$. MQC is defined as

$$MQC(P) = 2MQ(P) + |P| - Diff(P) - Iso(\mathrm{P}), \text{ where}$$

MQ (Modularization quality) of P is defined an adaptation of the MQ considered in [4]. In particular, let $\varepsilon_i$ be the intra cluster distance between the nodes in $\boldsymbol{C_i}$ defined as $\varepsilon_i = \sum_{v \in C_i, k \in C_i - \{v\}} weight(v, k)$. Let $\mu_i$ be the inter cluster distance for the cluster $\boldsymbol{C_i}$ defined as $\mu_i = \sum_{v \in C_i, k \in V - C_i} weight(v, k)$. For each $i \in 1 \cdots n$, let $CF_i$ be the *Clustering Factor* for the i^th Cluster $C_i$: $CF_i = \varepsilon_i / (\varepsilon_i + \mu_i)$. In terms of these, *MQ(P)* is defined as: $MQ(P) = \sum_{i=1}^{i=n} CF_i$ .

*Diff(P)* is the difference between the maximum and the minimum number of classes in a cluster in P. *Iso(P)* is the number of clusters in P consisting of only 1 or 2 classes. |P| is the number of clusters in P.

Step [3.1]: Generate initial Clustering:

(**SeedPopulation#1:** *Connected Components based*)
(1_i) Sort the list of edges in a non-increasing order.
(1_ii) Choose top quartile of the edges from the sorted list of step (1_i). Let the set of selected edges be $\mathbf{E_{topQ}}$.
(1_iii) Construct connected components using top weighted edges in $\mathbf{E_{topQ}}$ so long as both of the following conditions continue holding:
Condition1:: There are edges in $\mathbf{E_{topQ}}$, which are not used so far in any of the connected component.
Condition2:: Number of independent components are < 2*Number of Packages
Each connected component would give rise to a Cluster.
(1_iv) For each of the remaining nodes in $\boldsymbol{G_{ext_dep}}$ (i.e., which were not considered in step (1_iii)), associate it to that cluster, from which it has least distance.
Let $\mathbf{P_1}$ be the set of seed Clusters generated**.**

(**SeedPopulation#2:** *Inheritances based*)
(2_i) Sort the list of edges in a non-increasing order.
(2_ii) Define the set $\mathbf{E_{seed}}$ of seed edges for initial clustering as:
$E_{seed}$ = {Set of edges with non-zero inheritance similarity}
(2_iii) Identify connected components in the graph using only the selected edges in $\mathbf{E_{seed}}$. Each connected component would give rise to a Cluster.
(2_iv) For each of the remaining nodes in $\boldsymbol{G_{ext_dep}}$ (i.e., which were not considered in step (2_iii)), associate it to that cluster, from which it has least distance.
Let $\mathbf{P_2}$ be the set of seed Clusters generated**.**

(**SeedPopulation#3:** *Packaging based*)
(3_i) Choose each package as a seed cluster.
Let $\mathbf{P_3}$ be the set of seed Clusters generated**.**

(**SeedPopulation#4:** *Random Clustering*)
(4_i) Select a random number $n \in [1..d]$, where d is the number of classes.
(4_ii) For each class:
Generate a random number $k \in [1..n]$ and assign the class to the cluster $k$.
Let $\mathbf{P_4}$ be the set of seed Clusters generated**.**

(**SeedPopulation#5:** *k-mean based*)
(5_i) Select a random number $n \in [1..d]$, where d is the number of classes.
(5_ii) Generate initial clustering using k-means clustering technique [6] with parameter k = n.
Let $\mathbf{P_5}$ be the set of seed Clusters generated**.**

(**SeedPopulation#6:** *Clique based*)
(6_i) Sort the list of edges in a non-increasing order.

(6_ii) Choose top quartile of the edges from the sorted list of step (6_i). Let the set of selected edges be $\mathbf{E_{topQ}}$.

(6_iii) For each node, $v \in V$, calculate its clique strength as follows:

$$CliqueStrength[v] = \sum_{[(v,u)\in E_{topQ} OR\ (v,w)\in E_{topQ}]\ AND [(u,w)\in E_{topQ}]} w(u,w)$$

(6_iv) Sort the nodes based upon their clique strengths.

(6_v) Select the top quartile of the nodes as cluster centres.

(6_vi) For each of the remaining nodes, associated them to that cluster from which it has maximum overall similarity.

Let $\mathbf{P_6}$ be the set of seed Clusters generated**.**

Step [3.2]: (Test for clustering initiation)

*/* This test would check the modularity of the initial seed clusters and if it is already very high, return that as the clustering else would carry out clustering process by maximizing MQ. */*

[3.2.1] Estimate the density of the graph

$$density = \sum_{e\in E} w(e)$$

[3.2.2]

For each Seed Clustering $\mathbf{P_i}$ in Step[3.1]:

Let

$$MQC(P_i) = 2MQ(P_i) - \text{diff}(P_i) - \text{iso}(P_i)$$

Where

$$diff(P_i) = \text{maximum number of classes in any of the clusters} - \text{minimum number of classes in any of the clusters}$$

$$\text{iso}(P_i) = \text{number of clusters in } P_i \text{ containing only 1 or 2 claaases}$$

If $(MQC(P_i) > |P_i|)$ ($|P_i|$ is the number of clusters in $P_i$)

CloseToOptimal$_i$ = True

[3.2.3]

**If** (density > 0.5*|E|) & (CloseToOptimal$_1$ | … | CloseToOptimal$_5$)

Mark seed-clustering for which MQ is maximum.

Don't use this seed-clustering for further processing since it is already on an optimal level.

Goto Step[3.3].

**Else**

Goto Step[3.3]

Step [3.3]: Generate Partitions by Maximizing MQC

(using Hill Climbing + Simulated Annealing approach)

*Note: Variables* temp, α, mqOld, mqNew, *and* snTag *are Global.*

[3.3.1]

**For each** of the $\mathbf{P_1, P_2, \ldots, P_6}$, which were not marked in Step[3.2.3]:

*mqOld* = *-d*

*currentP* = $\mathbf{P_i}$

*temp* = 1000

/* Variable temp represents the temperature for simulated annealing function SN(). Default = 1000. (In general it should be of the order of the number of classes in the application source.) */

α = 0.7

/* Variable $\alpha\epsilon[0..1]$ is cooling constant used while applying the simulated annealing. Default = 0.7 */

*snTag = False*

*diff = maximum number of classes in any of the clusters in currentP – minimum number of classes in any of the clusters in currentP*

*iso = number of clusters in currentP consisting of only 1 or 2 classes*

*mqNew* =2*MQ(*currentP*) + |*currentP*| - *diff* - *iso*

**If** (*mqNew* -|*currentP*| < ∈) /*ϵ is a constant with very small value */ **return** *currentP*

**while** (*mqNew > mqOld) OR (snTag == True)*

*mqOld = mqNew*
*nextP = ClimbHill(currentP)*
*currentP = nextP*
*mqNew = MQ(currentP)*

**return** *currentP*

[3.3.2] Return the Clustering with highest MQC

*Note: An approach to identify and eliminate outlier type clusters is given in Appendix [B].*

**ClimbHill(…)**

Inputs: Clustering P = $\{C_1, C_2, \cdots, C_n\}$ .

Output**:** New clustering, which is a 1-neighbouhood of P, and will have an improved MQ.

*/* For a Clustering **P**, its 1-neighbour **nP** is defined as another Clustering, which is exactly same as **P** except that a single element (node) of a cluster in **P** is in a different cluster in **nP**. */*

Step [1c]

Clustering BestP = P
Diff[3] = {0, 0, 0}

*/* Diff array records the difference between a Clustering and its 1-neighbour. Diff[0] is the index of the node in V, which differs between the neighbours. Diff[1] is the index of the Cluster, where the node Diff[0] originally was and Diff[2] is the index of the cluster where node currently is.*/*

Counter = 0
improved = *False*
*snTag = False*
*mqBest = mqOld*

Step [2c]

[2.1c] **for** each node *u* in **V:**
Let $C$ be the cluster in which *u* is currently present.
Identify the cluster $C^u \neq C$, from which *u* has maximum similarity. $C^u$ is the target cluster for *u*.
[2.2c] For each pair of clusters $(C, C')$, estimate the number of nodes in $C$, for which $C'$ is the target cluster.
[2.3c] Sort the cluster pairs based upon the estimate in Step [2.2].

[2.4c] Based upon this list, identify that subset of nodes, which can be moved from their current cluster to the target cluster and would increase the value of MQ maximally. Return the clustering after this movement.
In case, when no such movement is possible, apply the following 1-neighbourhood movement with steepest hill climbing and simulated annealing:

NumOfNeigh = $|\boldsymbol{V}| \times |\boldsymbol{P}|$

/* (number of nodes in $G_{ext_dep}$) × (number of Clusters in P) */

Step [3c]

**while** Counter < NumOfNeigh

Clustering tempP = getNewNeigh(BestP, Diff[])
Count = Counter + 1
mqT = MQ_Rec(tempP, Diff[])
**if** (mqT > mqBest) **OR** SN(mqT, mqOld, temp)

/* Checks if neighbour tempP has an MQ better than any seen so far or if the simulated annealing function SN() allows a 1-neighbour of lower quality. If so, tempP is saved as BestP and the improved flag is set to true. */

BestP = tempP
mqBest = mqT
improved = **true**
**if** SN(mqT, mqOld, temp)
temp = α*temp
*snTag = True*
**return** BestP
**if** (count ≥ neigh#) **AND** (improved == true)
**return** BestP

Step [4c] **return** BestP

**SN(mqNew, mqOld, temp)**

Float θ = random()

/* A random number between 0 and 1 generated uniformly */

if ($mq_{new} < mq_{old}$) AND (θ < $e^{\frac{mq_{new}-mq_{old}}{temp}}$)
return **true**
return **false**

**MQC(…)**

Inputs**:** Clustering $\{C_1, C_2, \cdots, C_n\}$.

Output: MQ for the given clustering.

Procedure**:**

Step [1] $\varepsilon_i$ is the intra cluster distance for the cluster $\boldsymbol{C_i}$ calculated as

$$\varepsilon_i = \sum_{v \in C_i, k \in V - C_i} weight(v, k)$$

Step [2] $\mu_i$ is the inter cluster distance between the cluster $\boldsymbol{C_i}$ calculated as

$$\mu_i = \sum_{v \in C_i, k \in C_i - \{v\}} weight(v, k)$$

Step [3] Modularisation Quality (MQ) of input Clustering is calculated as
[3.1] For each $i \in 1 \cdots n$

$$CF_i = \frac{\varepsilon_i}{\varepsilon_i + \mu_i}$$

$CF_i$ is the *Clustering Factor* for the i$^{th}$ Cluster $C_i$.
[3.2] Calculate MQ as:

$$MQ = \sum_{i=1}^{i=n} CF_i$$

Step [4] Calculate MQC similar to Step [3.2.2]
Step [5] Return MQC.

**MQ_Recur(…)**

Inputs**:** Clustering $\{C_1, C_2, \cdots, C_n\}$, mqOld, node $k$, Cluster $C_i$, and Cluster $C_j$.

Output**:** MQ for the new clustering.

Procedure**:**

Note: After movement $C_i = C_i - \{k\}$ ***and*** $C_j = C_j + \{k\}$
Recalculate MQ as follows:
$MQ_{updated}$ =

$$MQ - \left(CF_i + CF_j\right) + \frac{\varepsilon_i - \Delta_{k,i}}{\varepsilon_i + \mu_i - (\Delta_{k,i} + \sum_{m \neq i}^{1 \leq m \leq n} \delta_{k,m})} + \frac{\varepsilon_j + \Delta_{k,j}}{\varepsilon_j + \mu_j + (\Delta_{k,j} + \sum_{m \neq j}^{1 \leq m \leq n} \delta_{k,m})}$$

Where
$CF_r$ is the original Clustering Factor for the r$^{th}$ Cluster $\boldsymbol{C_r}$

$\Delta_{k,i} = \sum_{v \in C_i} weight(k, v)$

$$\Delta_{k,j} = \sum_{v \in C_j} weight(k, v)$$

$$\delta_{k,m} = \sum_{v \in C_m} weight(k, v)$$

$\varepsilon_i$ is the (original) inter cluster distance for the cluster $\boldsymbol{C_i}$**.** Same for $\varepsilon_j$

$\mu_{i,m}$ is the original intra cluster distance between the cluster $C_i$ ***and*** $C_m$. Same for $\mu_{j,m}$.

# Step [4]: Cluster Interfaces

Using the clusters generated in Step [3] and the source code of the application, identify (public) methods of all the classes in each cluster, which are being called by the classes in other clusters. List of such methods will constitute the interfaces of each cluster.

## Step [5]: Inter Cluster Interactions

Inputs: Clusters from Step [3] and the dependency graph from Step[1.5].

Output: Inter Cluster Interaction Graph in GraphML format.

Procedure:

Step[5.1]: For each Cluster pair (Cluster N, Cluster M), generate the list of (public) methods of Cluster N, which are being called by the classes in Cluster M. This list will define the inbound dependencies between Cluster N and M.
Step[5.2]: Generate the inbound dependency Graph for the Clusters as follows:
[5.2.1]: For each Cluster, there is a node with Cluster number as a node id.
[5.2.2]: If the list of inbound dependencies between Cluster N and M as generated in Step [5.1] is not empty, there will be a directed edge from node with id N to node with id M, with attribute as the list of methods in the list.

## Step [6]: Cluster Auto-labelling

For each cluster:

**(Approach#1)**

Step[6.1]: Select the most frequent (i.e., having high tf-idf values) concept words appearing the class names for the classes contained in the component. Default choice: top 5.
Step[6.2]: Label the component in terms of these concept words signifying the dominant functions implemented by the component.
Step[6.3]: Identify the centroid of the cluster and mark it in the component diagram as the centre of the component.

**(Approach#2)**

Step[6.1]: Apply LDA (Latent Dirichlet Allocation) [8] to extract top 5 dominant terms using the class-names, textual vectors, and public method identifiers.

## Step [7]: Visualizing Clusters and their Interactions

Step[7.1]: Visualize each cluster generated in Step [3] as a polygon with nodes corresponding to classes in the Cluster and edges between each pair of classes having strength corresponding to the combined similarity between the corresponding classes as estimated in the Step[2].

1. Classify combined similarity into 5 groups: (0, 0.2], (0.2, 0.4], (0.4, 0.6], (0.6, 0.8], and (0.8, 1].
2. For each group select different edge type with different edge thickness.

Step[7.2]: Using standard graph visualization techniques, display the inter component interactions discovered in the Step[5].

Step[7.3]: Display the interaction between the discovered Clusters and the classes scoped out in Step[0] using dependency graph of the application, i.e., for each component visualize the UI, Models, Utils, DA objects used.

## Step [8]: Borderline Classes

Step[8.1]: Identify borderline classes. A borderline class is one which has high similarity with a class from some other cluster. Identify these classes and display their relationship in component diagram.
Step[8.2]: Allow user to manually reassign these borderline classes to other clusters.
Step[8.3] Post user reconfiguration, perform Step[4] to Step[7] to re-calculate and visualize updated Clusters, their interfaces, and their interactions.

## Step [9]: Mapping Functional Entities to Clusters

Additional Inputs: User provides a naïve functional model consisting of only functional entity descriptions. The descriptions can be single word names or more elaborate textual descriptions.

Procedure:

Step[9.1]: (Converting each functional entity (name) into a word vector)
[9.1.1] Split the functional entity into distinct words.
[9.1.2] In terms of these words, represent the functional entity as a vector analogous to a document vector in the co-occurrence matrix C (ref. Step[2.1.1]).
[9.1.3] (Applying tf-idf)
For each of the words in the vector,
If the word also appears in the list of previously extracted IR tokens in Step[2.1.1(ii)], apply the idf same as was earlier calculated in the context of the co-occurrence matrix C.
Else if the word only appears in the functional entity name, apply its idf factor as $\ln N$, where $N$, is the number of classes in the application.
Step[9.2]: For each Cluster
[9.2.1]: Generate cluster level class name concept vectors by summing the class level text vectors (generated in Step[2.2.2]) for the classes constituting the Cluster.
[9.2.2] Generate cluster level sets of the class names by summing the class level vectors of word concepts appearing in the class names (generated in Step[]) for the classes constituting the Cluster.
Step[9.3]: Calculate similarity between each of the functional entities and clusters using the text and the class names concept vectors generated in Step[9.1] and Step[9.2].
Step[9.4]: Rank the clusters in the decreasing order of their distances from each of the functional entities.
Step[9.5]: For each entity, select the clusters having similarity more than a minimum threshold.
Step[9.6]: Visualize and report the functional entity to component mapping.

## Step [10]: (Entitiy2Code Mapping) Mapping Queries to Classes

Additional Inputs: User provides a query as a text string consisting of one or more words.

Procedure:

Step[10.1]: Similar to Step[9.1], convert the query string into a word vector and apply tf-idf on the vector.

Step[10.2]: Calculate similarity between the query vector and each of the classes using the text vector and the class names concept vectors. Let $\delta_{VSM}[query, i]$
Step[10.3]: Rank the classes in the decreasing order of their averaged distances from the query vector in terms of text vectors and the class name vectors.
Step[10.4]: Select top r classes from the list. Default r = d (number of classes in the application). Let the list of these top r classes be VSMRankList.
Step[10.5]: For each class with rank $i \in [1..r]$ in the VSMRankList:
[10.5.1]: Compute its total ranked combined similarity with all other classes using the combined similarity similar to the page-ranking algorithm
Step[10.6]: Rank each class in VSMRankList based upon its similarity value computed in step [11.5]. Let the ranked list be CentroidRankList.
Step[10.7]: For each class in VSMRankList:
Final rank of a class =
α.(Rank in VSMRankList) + β.(Rank in CentroidRankList)
where α,β ϵ [0,1]; α+ β=1. Initially we can choose α = 0.6 and β = 0.4.
Step[10.8]: Display top 5 classes as answers based upon the final ranks in Step [10.7].

# Step [11]: Cluster Hierarchy

After the first level of Clustering in Step [3], each cluster is considered as an atomic code object and these code-objects are clustered together using the same partitioning based clustering using the maximization of the MQC (modularization quality constraint) as the selection criteria as applied in Step[3].
The similarity between two clusters is calculated as the normalized summation of the similarity scores between the pairs of entities (e.g., classes for the first level of clusters) across the clusters. Formally,
Let two clusters $CL_1 = \{c_{11}, c_{12}, \cdots, c_{1k}\}$ and $CL_2 = \{c_{21}, c_{22}, \cdots, c_{2r}\}$ be the output of previous level of clustering.
Similarity between $CL_1 and\ CL_2$ is calculated as

$$sim(CL_1, CL_2) = \frac{1}{k * r} \sum_{(c_{1j}, c_{2l}) \in CL_1 \times CL_2} sim(c_{1j}, c_{2l})$$

This process is repeated to higher levels of hierarchy until there is single cluster or MQ cannot increase any further since similarities between any two code objects is relatively very low (close to zero).

# Step [12]: Clustering Application Portfolios

Objectives:

1. Create Clustering of classes within each of these applications separately and
2. Generate Clustering only among applications.

Procedure:

Step[12.0]: For (i), apply Steps [1] to [3] to each application separately.
Step[12.1] (Estimate Similarities between Applications)
[12.1.1] (Textual Similarity):
(i) Create a giant text vector for each application as a multi-set union of the vectors for the classes in it
(ii) Apply TF-IDF by recalculating the IDF factor across all classes appearing in all applications.
(iii) Calculate cosine similarity for each pair of applications using these giant text vectors (similar to Step[2.1.3]).

[12.1.2] (Class Name Similarity):
(i) Similar to [12.1.1](i).
(ii) Similar to [12.1.1](ii).
(iii) Estimate similarity for each pair of application similar to Step[2.2.3].
[12.1.3] (Dependency Based Similarity): Identify the cross component dependencies and then follow all the steps similar to steps [2.7.1] to [2.7.5].
[12.1.4] (Combined Similarity): Similar to Step[2.8] - linear combination of similarities estimated in the Step[12.1.1]-[12.1.3].
Step[12.2] (Application Clustering) Using the combined similarity from step[12.1.4] apply same Step[3] for clustering.

# APPENDIX [A]: AN APPROACH TO WEIGHT ASSIGNMENT PROCESS

The choice of the significance factors has critical bearing on the combined similarity scores and in turn on the generated clusters. It is not clear whether there exist certain choices, which might in general give better results than others. A rule of thumb, however, in absence of domain knowledge could be that, if all the features are well represented in the code, equal treatment of all these features is a safe choice.

Because quality of semantic features is usually dependent upon the maturity of the application development process, we need to be careful in cases, where a feature has high variance in its distribution across classes in the application. For example, cases where some of the classes are well documented while many others having no documentation at all or some class-names are much more informative, while others are poorly named. In these situations, it is advisable to set the corresponding SFs for these features lower compared to others so that to avoid the problem of non-uniformly measuring combined similarity across different class pairs. On the other hand, for the syntactic features, it is not possible to make any assessment of their quality in absence of user given inputs or meta-information.

# APPENDIX [B]: IDENTIFICATION AND ELIMINATION OF OUTLIER TYPE CLUSTERS

**Identification of Outlier type Clusters::**

Let C1, C2, .. Cn be the clusters generated using CC based technique with top edge weight threshold set to α percentile.

[Step1] Estimate *coefficient of  skewness* of the cluster size list. Skewness is the measure of asymmetry in data set. There are several formulations to estimate skewness of a data set. We consider the unbiased Fisher's Measure of Skewness [8]. For this following formula can be used:

$$G_1 = \frac{n}{(n-1)(n-2)}\left(\left[\sum_{i=1}^{n}(x_i-\bar{x})^3\right]\Big/\left[\frac{1}{n-1}\sum_{i=1}^{n}(x_i-\bar{x})^2\right]^{3/2}\right)$$

$x_i$ is the size of the ith cluster Ci and $\bar{x}$ is the mean cluster size.

**<u>Elimination of Outlier type Clusters::</u>**

[Step2]
**If** ($G_1$ > 2):
Tag each cluster as OUTLIER having size > size of the median cluster.

[Step3]
**If** the list of OUTLIER clusters is not empty:
**If** ($\alpha < 99$)
// (Increase $\alpha$)
**If** $\alpha < 95$
Use binary search type increment i.e., $\alpha = \alpha + [(99 - \alpha)/2]$
**Else**
$\alpha = \alpha + 1$
Apply connected component algorithm with new $\alpha$ and generate clusters.
**Else**
Apply min-cut on the OUTLIER clusters to break these into sub-clusters.
[Step4] Repeat Step[1] to Step[3] if necessary.

# REFERENCES

[1] A. Wiese, V. Ho, and E. Hill, “A Comparison of Stemmers on Source Code Identifiers for Software Search,” Proc. 27th IEEE International Conference on Software Maintenance (ICSM’11), IEEE, pp. 496–499, 2011.

[2] C.-H. Lung, M. Zaman, and A. Nandi, “Applications Of Clustering Techniques to Software Partitioning, Recovery and Restructuring,” J. Syst. Softw., vol. 73(2), pp. 227-244, 2004.

[3] Michel M. Deza and Elena Deza, “Encyclopedia of Distances,” Springer, 1st edition, 2009.

[4] Brian S. Mitchell and S. Mancoridis, “On the Automatic Modularization of Software Systems Using the Bunch Tool,” IEEE Trans. Softw. Eng., vol. 32(3), pp. 193-208, 2006.

[5] K. Praditwong, M. Harman, and X. Yao, “Software Module Clustering as a Multi-Objective Search Problem,” IEEE Trans. Softw. Eng., vol. 37(2), pp. 264-282, 2011.

[6] J. A. Hartigan and M. A. Wong, “Algorithm AS 136: A k-means Clustering Algorithm,” Journal of the Royal Statistical Society. Series C (Applied Statistics), vol. 28(1), pp. 100–108, 1979.

[7] David M. Blei, Andrew Y. Ng, and Michael I. Jordan, “Latent dirichlet allocation,” J. Mach. Learn. Res., vol. 3, Mar. 2003, pp. 993-1022.

[8] Doane, D. P. and Seward, L. E. "Measuring Skewness: A Forgotten Statistic?" Journal of Statistics Education, Vol. 19(2), 2011. Avialable online at http://www.amstat.org/publications/jse/v19n2/doane.pdf.

[9] Niels Streekmann, “Clustering-Based Support for Software Architecture Restructuring,” Springer, 2011.

[10] P. Andritsos and V. Tzerpos, “Information-Theoretic Software Clustering,” IEEE Trans. Softw. Eng., vol. 31(2), pp. 150-165, 2005.